# ChatGPT Is More Likely to Be Perceived as Male Than Female

Jared Wong [1, a] and Jin Kim [1, 2, a]

[1] Department of Marketing, Yale School of Management

[2] Advanced Institute of Business, School of Economics and Management, Tongji University

## Author Note

Jared Wong  https://orcid.org/0000-0001-8269-5579

Jin Kim  https://orcid.org/0000-0002-5013-3958

Correspondence concerning this article should be addressed to Jin Kim (Advanced Institute of Business, Tongji University, Shanghai, China. Email: jin.m.kim@yale.edu).

**Abstract**

We investigate how people perceive ChatGPT, and, in particular, how they assign human-like attributes such as gender to the chatbot. Across five pre-registered studies ($N = 1{,}552$), we find that people are more likely to perceive ChatGPT to be male than female. Specifically, people perceive male gender identity (1) following demonstrations of ChatGPT's core abilities (e.g., providing information or summarizing text), (2) in the absence of such demonstrations, and (3) across different methods of eliciting perceived gender (using various scales and asking to name ChatGPT). Moreover, we find that this seemingly default perception of ChatGPT as male can reverse when ChatGPT's feminine-coded abilities are highlighted (e.g., providing emotional support for a user).

*Keywords:* ChatGPT, perception of gender, artificial intelligence, chatbot, technological innovation

**ChatGPT Is More Likely to Be Perceived as Male Than Female**

OpenAI's ChatGPT (Generative Pretrained Transformer) is one of the most advanced large language models to date (Rudolph et al., 2023), carrying wide-ranging implications for public health (Biswas, 2023b), education, (Kung et al., 2023), translation (Jiao et al., 2023), writing (Kitamura, 2023), and scientific research (van Dis et al., 2023). In its own words:

> As an AI language model, the goal of ChatGPT is to generate human-like responses to natural language input from users, in order to provide helpful and informative conversation, as well as assistance in tasks such as answering questions, making recommendations, or engaging in small talk (OpenAI, 2023).

Trained with reinforcement learning from human feedback (RLHF), this language model allows users to engage in naturalistic conversation. Concretely, users can ask ChatGPT, for instance, to write wedding vows or religious sermons (Kelly, 2023; Willingham, 2023), re-write the first scene of *Death of a Salesman* using characters from *Frozen* (Thorp, 2023), or aid in the process of research (e.g., data analysis and interpretation, scenario generation, and abstract writing) (Biswas, 2023a; Gao et al., 2022). Given the potential utility for both academic and non-academic pursuits, ChatGPT has become a "cultural phenomenon" (Thorp, 2023, p. 313), with lay audiences learning about this technology from the popular press (e.g., the Wall Street Journal, the New York Times, and CNN). These articles answer questions like "How Should I Use A.I. Chatbots Like ChatGPT?" and "What Is ChatGPT?", demonstrating the extent to which ChatGPT has entered the public discourse. Indeed, lay audiences are beginning to think about the role of AI chatbots in parenting (Caron, 2023), privacy (Satariano, 2023), and the relationship between the mind and body (Whang, 2023).

In contrast to prior manifestations of artificial intelligence (AI), ChatGPT stands out as an exception because of its ability to engage in sophisticated and naturalistic conversation (Lund & Wang, 2023). This ability to communicate like humans invites potential assignment of human-like qualities. Indeed, a large body of research suggests that people spontaneously anthropomorphize non-human entities (Heberlein & Adolphs, 2004). For instance, people perceive a sense of animacy and intentionality from moving shapes (Gao et al., 2019; Heider & Simmel, 1944), gender from point-light displays (Kozlowski & Cutting, 1977), and biological motion from dots (Johansson, 1973). In a similar manner, people may anthropomorphize ChatGPT and perceive it to embody human-like attributes.

Past investigations have considered the identities of technological assistants, such as Amazon's Alexa and Apple's Siri, who have gendered names and voices (Kepuska & Bohouta, 2018; Loideain & Adams, 2020). Specifically, scholars have shown that people percieve Alexa to be female (Fortunati et al., 2022), while others remain critical of the perpetuation of gendered stereotypes by arguing that such assistants represent "digitial domesticity" (Woods, 2018). Such research suggsts that technological assistants, due to their roles being coded as feminine, are largely percieved to be female in the eyes of users. If a similar logic is applied, then we may expect ChatGPT to be perceived as female, as it can also serve as a technological assistant.

Despite this prediction from the extant literature, a gender stereotype account might predict the opposite finding. Research on gender stereotypes find that women and men are perceived differently along the dimensions of warmth and competence, whereby women are associated more with warmth and men more with competence (Ebert et al., 2014; Fiske et al., 1999; Heflick et al., 2011; Otterbacher et al., 2017; Ramos et al., 2018). To the extent that such biased gender stereotypes are embedded in people's minds, the capabilities of ChatGPT (i.e., its competence) may lead people to perceive it to be more male than female.[1] However, the perception of ChatGPT as male can reverse if its features signaling warmth are made more salient than those signaling competence. We test these two predictions below.

## Study 1: Is ChatGPT More Likely to Be Perceived as Male?

**Overview and Methods**

Study 1 serves as our initial attempt to uncover a potentially biased perception of ChatGPT's putative gender (i.e., the tendency to perceive the non-human chatbot to be more male than female). We recruited 501 participants ($M_{age}$ = 38, 50% male, 48% female, 2% other) from Prolific. All watched a brief video clip (around one-minute long) depicting how a user might interact with ChatGPT. These videos illustrated how a user could ask ChatGPT to (1) provide information, (2) generate text, (3) assist with coding, or (4) summarize text, which form our operationalization of ChatGPT's core capabilities.[2] These stimuli were randomly assigned to participants between-subjects and served as an attempt to sample stimuli (Wells & Windschitl, 1999). Participants then reported whether they perceived ChatGPT to be more male or female on an 8-point scale (1 = *Definitely female*; 8 = *Definitely male*) or in a binary choice (male vs. female).[3] These scales were randomly assigned between-subjects and the order of answer choices (e.g., whether the male or female option was presented first) was counterbalanced.

---

[1] We use the term *perception of gender* to refer to judgments about ChatGPT being more male or female, while acknowledging that gender is no longer considered binary but rather a spectrum (Richards et al., 2016). We refrain from *perception of sex* given that ChatGPT does not have a biological sex. Also, it is important to note that despite our usage of the word "perception," we are *not* suggesting that there are male or female *percepts* in the context of ChatGPT (in the same way there would be for bottom-up processes; Firestone & Scholl, 2016).

[2] The video clips showed conversations between ChatGPT and a user about (1) theory of evolution (in the Providing Information condition), (2) drafting an email to a Prolific study participant (in the Generating Text condition), (3) how to simulate random events using the programming language Python (in the Coding Assistance condition), and (4) summarizing a news article on a recent national security event (in the Summarizing Text condition). The video clips will be available on the project's OSF page: https://osf.io/

[3] Here, we used an even-numbered scale rather than an odd-numbered scale, because using the odd-numbered scale would allow people who might have a very subtle gender association to mask this association in their minds by choosing the scale midpoint. However, for robustness, we use odd-numbered scales in other studies.

**Results and Discussion**

Following our pre-registered analysis plan, we collapsed data across the four ChatGPT capability conditions and examined participants' response on each of the two response modes (binary choice and 8-point scale). Among participants assigned to respond to the binary choice, 75% (95% CI = [69%, 80%]) perceived ChatGPT to be male. Among those assigned to respond on the 8-point scale, the mean perceived gender rating was significantly different from the midpoint of the scale (4.5) in the direction of male ($M = 4.00$, $SD = 1.19$), $t(250) = 6.68$, $p < .001$, $d = 0.42$.

If we deviate from our pre-registered analysis plan and examine the perceived gender responses *within each* of the four ChatGPT capability conditions, we find the same pattern of results. Regardless of the four ChatGPT capabilities to which they were exposed, participants were more likely to perceive ChatGPT to be male than female on the binary choice (all four $p$s < .01). Similarly, on the 8-point scale, the mean perceived gender rating was lower than the scale midpoint in the male direction, significantly so in three of the four ChatGPT capability conditions (providing information, assisting with coding, and summarizing text; all $p$s < .008) and directionally so in one of the four (generating text; $p = .13$). These results show that people perceive ChatGPT to be more male than female (1) on different response scales (binary choice or an 8-point scale) and (2) after seeing various capabilities of ChatGPT.

## Study 2: Robustness Check

**Overview and Methods**

To probe the robustness of the perception of ChatGPT as more male than female, we varied the video stimulus and response scale. In Study 2, we recruited 201 participants from Prolific and exposed all participants ($M_{age}$ = 39, 51% male, 48% female, 1% other) to a different instantiation of ChatGPT providing information for a user (as in one of the four conditions in Study 1).[4] We then randomly assigned participants to either a counterbalanced trinary choice condition (Male vs. Neutral vs. Female) or 7-point scale (1 = *Definitely male*; 7 = *Definitely female*) condition.

**Results and Discussion**

In Study 2, most participants responding on the ternary choice of perceived gender (93%) responded with the "Neutral" choice. However, the choice share of "Male" (7%) was significantly greater than the choice share of "Female" (0%), $p$ < .008.[5] For participants responding on the 7-point scale, participants' mean perceived gender rating was significantly different from the scale midpoint (i.e., 4) in the "Male" direction ($M$ = 3.90, $SD$ = 0.60) in a pre-registered one-tailed *t*-test, $t(99)$ = 1.68, $p$ < .05, $d$ = 0.17.

## Studies 3-4: Naming ChatGPT

**Overview and Methods**

In Studies 3 and 4, we used an implicit method to examine people's perception of ChatGPT's gender. Instead of explicitly asking participants to report which gender they perceived ChatGPT to be, we asked participants to give ChatGPT a human first name. We then displayed this name to the participants on the next page and asked them to indicate the gender most associated with the name.

---

[4] We varied the video stimuli to ensure that our finding was not driven by an artifact of the specific content. In this stimulus, we have participants watch how ChatGPT can provide information about Hegelian philosophy (rather than about theory of evolution as in Study 1).

[5] We slightly deviated from the pre-registered analysis for expositional clarity. However, following our original analysis plan (i.e., one-sample *t*-test) resulted in the same qualitative finding That is, the mean perceived gender rating ($M$ = 1.93, $SD$ = 0.26) significantly differed from the midpoint (2) of the 3-point scale (Male vs. Neutral vs. Female) in the "Male" direction, $p$ = .004, $d$ = 0.27.

Study 3 sought to examine people's spontaneous perception of ChatGPT's gender *in the absence* of any substantive information provided by the researchers. A total of 400 participants ($M_{age}$ = 40, 50% male, 48% female, 2% other) were recruited from Prolific to participate in Study 3. They began the study by learning that the researchers wanted to learn about the participants' "thoughts on ChatGPT, an artificial-intelligence chatbot that can hold conversations with humans." This description of ChatGPT was the only piece of information we provided. Participants were then asked, "If you were to give ChatGPT a human name (first name only), what name would you give it?" After entering a name for ChatGPT, participants proceeded to the next page, which displayed the name they entered: "You said you would give ChatGPT the following name: '[the name entered by the participant].'" Participants were then asked to indicate the gender of the name they entered: "In your opinion, which gender is most associated with the name '[the name entered by the participant]'? (Male / Neither Male Nor Female / Female)." Afterwards, participants reported whether they had heard of ChatGPT before participating in the study and whether they had used it themselves. They provided demographic information to complete the study.

Study 4 was similar to Study 3 in its procedure and analysis plan, except that it showed all participants a video clip of how ChatGPT can be used (i.e., summarizing text, as in Study 1) before they gave ChatGPT a name. By demonstrating ChatGPT's interaction with a user, we were able to ensure that all participants had some idea of how ChatGPT worked. A total of 150 participants ($M_{age}$ = 40, 50% male, 49% female, 1% other) were recruited from Prolific for Study

4. Participants were told at the beginning that they would "learn about ChatGPT" and proceeded to the next page, where they viewed a 41-second video clip of a user interacting with ChatGPT, in which ChatGPT summarized a news article for the user through specific instructions in a dialogue. Afterwards, the procedure was identical to that of Study 3: Participants gave a name to ChatGPT, indicated the gender of the name that they just gave, and completed exploratory measures (knowledge and prior usage of ChatGPT), along with demographic measures.

**Results and Discussion**

In Study 3, we followed our pre-registered analysis plan to exclude names that participants categorized as "Neither Male Nor Female" (15% of the 400 names; $n = 59$). Then, with the remaining names that participants categorized as either "Male" or "Female" ($n = 341$), we conducted a binomial test. Of these, 81% ($n = 275$) were categorized as male (95% CI = [76%, 85%]), which was significantly greater than 50%, $p < .001$. As hypothesized, participants were more likely to give ChatGPT a male name than a female name. Thus, even in the absence of any substantive information provided by researchers, participants seemed to perceive ChatGPT as more male than female.

We also conducted the same binomial test above after determining the names' gender more objectively using an R package "gender" (Mullen, 2021).[6] One concern with analyzing the gender that participants themselves indicated for their name is that participants may have indicated gender incorrectly. For example, a participant might give ChatGPT the name "Alex," without considering the name's gender. Subsequently, however, when prompted to indicate the name's gender, the participant might indicate that the name is actually associated with neither

---

[6] When inferring the gender of names using the R package "gender," we used the package's default settings. This meant that the gender inference was based on the U.S. Social Security Administration baby name data from 1932 to 2012.

gender—so as to avoid being perceived as having a male-favoring bias in naming ChatGPT. Alternatively, a participant might give ChatGPT the name "Max" thinking it was a female name (as in the TV show *Stranger Things*), even though most people named Max in reality are male. For the latter kinds of unintentionally incorrect gender indications, it would make more sense to analyze the gender that participants indicated themselves (as reported earlier) because asking them to name ChatGPT was merely an *implicit* way of examining the gender they perceived. However, for the former kinds of intentionally incorrect gender indications, the objective inference of gender using the R package would be more appropriate because doing so would bypass any misrepresentation in the perceived gender response resulting from social desirability concerns.

Analyzing gender determined by the R package led to the same conclusion. Of the 400 names, the "gender" package could not determine the gender of 51 names (13%; e.g., "Chatbro"). However, of the remaining names ($n = 349$; 87%) whose gender the package could infer, 83% were male (95% CI = [78%, 87%]), which was significantly greater than 50%, $p < .001$. In short, the objective method of determining gender of the names also showed that ChatGPT was more likely to be perceived as male than female.

In Study 4, we again followed the pre-registered analysis plan and excluded names that participants categorized as "Neither Male Nor Female" (15% of the 150 names; $n = 23$). We then conducted a binomial test with the remaining names that participants categorized as "Male" or "Female" ($n = 127$). Of these, 85% ($n = 108$) were categorized as male (95% CI = [77%, 91%]), which was significantly greater than 50%, $p < .001$. As hypothesized, participants were more likely to give ChatGPT a male name than a female name, again implying that ChatGPT was more likely to be perceived as male than female.

Moreover, analyzing gender determined by the R package "gender" led to the same conclusion. Of the 150 names, the "gender" package could not determine the gender of 24 names (16%; e.g., "chappy"). However, of the remaining names ($n = 126$; 84%) whose gender the package could infer, 84% were male (95% CI = [76%, 90%]), which was significantly greater than 50%, $p < .001$. Objective determination of gender of the names provided further support to the hypothesis that ChatGPT was more likely to be perceived as male than female. All names given to ChatGPT by two or more participants in Studies 3 and 4 are listed in Table A1 in the Appendix.

### *Repeating the Analyses After Removing Names That Sound Similar to ChatGPT*

In both Studies 3 and 4, the most popular name was "Chad" (given to ChatGPT by 7-8% of participants). Participants may have named ChatGPT "Chad" because the name sounds similar to "ChatGPT." If names that sound similar to ChatGPT were more likely to be male names (e.g., Chad or Charlie), then prevalence of such names could explain the results that ChatGPT was more likely to be perceived as male than female in our naming-ChatGPT paradigm. To test this account, we repeated the analyses above after excluding all the names that began with "Ch" (including all the possible capitalization patterns of the two letters).

Results remained the same. In Study 3, excluding 73 names that begin with "Ch/ch/CH/cH" left us with a set of 268 names that participants perceived to be either male or female. A binomial test revealed that 78% ($n = 209$) of the 268 names were categorized as male (95% CI = [72%, 83%]), which was significantly greater than 50%, $p < .001$. Similarly, in Study 4, excluding names that begin with "Ch/ch/CH/cH" left us with a set of 103 names that participants perceived to be either male or female. A binomial test revealed that 83% ($n = 86$) of the 103 names were categorized as male (95% CI = [75%, 90%]), which was significantly greater than 50%, $p < .001$.

## Study 5: Reversing the Male Perception

**Overview and Methods**

We conjecture that the perception of ChatGPT as more male than female may stem from the capabilities demonstrated by ChatGPT and the stereotypical association between male and competence (Eagly & Mladinic, 1994). If this account of gender stereotypes underlies the biased male perception we find, then we should expect a reversal of the male perception when ChatGPT's capabilities signal warmth more than competence. In Study 5, we include a new instantiation of ChatGPT's capabilities to operationalize warmth: We presented a video clip of ChatGPT providing emotional support for a user, as if it were a therapist. In addition, we operationalized competence by presenting the video clip of ChatGPT summarizing text (as in Studies 1 and 4). Thus, we randomly assigned 300 participants ($M_{age}$ = 39, 50% male, 48% female, 1% other) to one of two conditions: Emotional Support vs. Summarizing Text. Then, all participants answered the counterbalanced 8-point scale used in Study 1 (1 = *Definitely female*; 8 = *Definitely male*).

**Results and Discussion**

Participants assigned to the Emotional Support condition perceived ChatGPT to be more female ($M$ = 4.83, $SD$ = 1.15) than those assigned to the Summarizing Text condition ($M$ = 4.13, $SD$ = 1.35), $t(298) = 4.85$, $p < .001$, $d = 0.56$. Moreover, replicating previous results, the mean perceived gender rating in the Summarizing Text condition was significantly different from the scale's midpoint (4.5) in the "Male" direction ($M$ = 4.13, $SD$ = 1.35), $t(150) = 3.36$, $p < .001$, $d = 0.27$. Importantly, the Emotional Support condition also resulted in the mean perceived gender rating significantly different from the scale's midpoint (4.5) but in the "Female" direction ($M = 4.83$, $SD = 1.15$), $t(148) = 3.54$, $p < .001$, $d = 0.29$. Thus, by highlighting a ChatGPT's capability that signaled warmth (i.e., providing emotional support like a therapist), we found the focal result in reverse, with participants now perceiving ChatGPT to be more female than male.

## Internal Meta-Analysis

To gauge the prevalence of male perception in ChatGPT, we conducted an internal meta-analysis of all the studies we conducted. Because we include all our studies in this analysis (i.e., "an empty file drawer"), and because every study was pre-registered to answer the same question ("Do people perceive ChatGPT to be more male or female?"), the internal meta-analysis can permit a valid inference of its results (Vosgerau et al., 2019), especially with the caveat that the results should be treated as exploratory rather than confirmatory.

We analyzed perceived gender of ChatGPT in each of the smallest subgroups within each study based on study design. For example, in Study 1, participants were randomly assigned to view one of the four capabilities of ChatGPT and were further randomly assigned to report the perceived gender either in a binary choice or on an 8-point scale. Thus, there were 8 (4 ChatGPT capabilities × 2 response modes) subgroups of participants based on study design, and within each subgroup, we estimated the percentage of answers that were "Male" (binary choice) or leaning toward male (e.g., the four out of the eight scale points on the male side). For Studies 2-4 where participants could respond with neutral answers (e.g., "Neutral," "Neither Male Nor Female," or the scale midpoint in on a 7-point scale), we excluded the neutral answers from our analyses following the pre-registered analysis plan. As a result, the percentage perceiving ChatGPT to be male in each subgroup of participants was estimated only among the participants who leaned toward either gender (male or female).

The results are shown in Figure 1. Within 13 of the 14 subgroups analyzed, the percentage of participants perceiving gender of ChatGPT to be male—among those who leaned toward either male or female—was greater than 50%. Moreover, 11 of these 13 percentages were significantly greater than 50% as their 95% CI did not cross 50%. However, among participants who viewed ChatGPT providing emotional support for a user (which would traditionally be considered a feminine-coded capability), the percentage of participants perceiving ChatGPT to be male (31%) was significantly less than 50%.

Across the 14 participant subgroups examined, a majority of participants who perceived (or leaned toward perceiving) gender of ChatGPT to be either male or female ($N = 1,289$) perceived it to be male ($n = 915$; 71%). The weighted mean percentage (of those perceiving ChatGPT as male) estimated using the generalized linear mixed model (GLMM) method and random effects approach was 74%, with its 95% CI = [65%, 81%] excluding 50% (Kim 2023; Schwarzer & Rücker, 2022; Viechtbauer, 2010). Thus, across all our studies—where we manipulated salience of ChatGPT's different capabilities, amount of information on ChatGPT (minimal vs. video stimuli), and elicitation method (e.g., various scales or asking to name ChatGPT)—participants were more likely to perceive ChatGPT to be male than female.

**Figure 1**

*Forest Plot of Percentages of Participants Perceiving ChatGPT to Be Male (Among Those Perceiving It to Be Male or Female)*

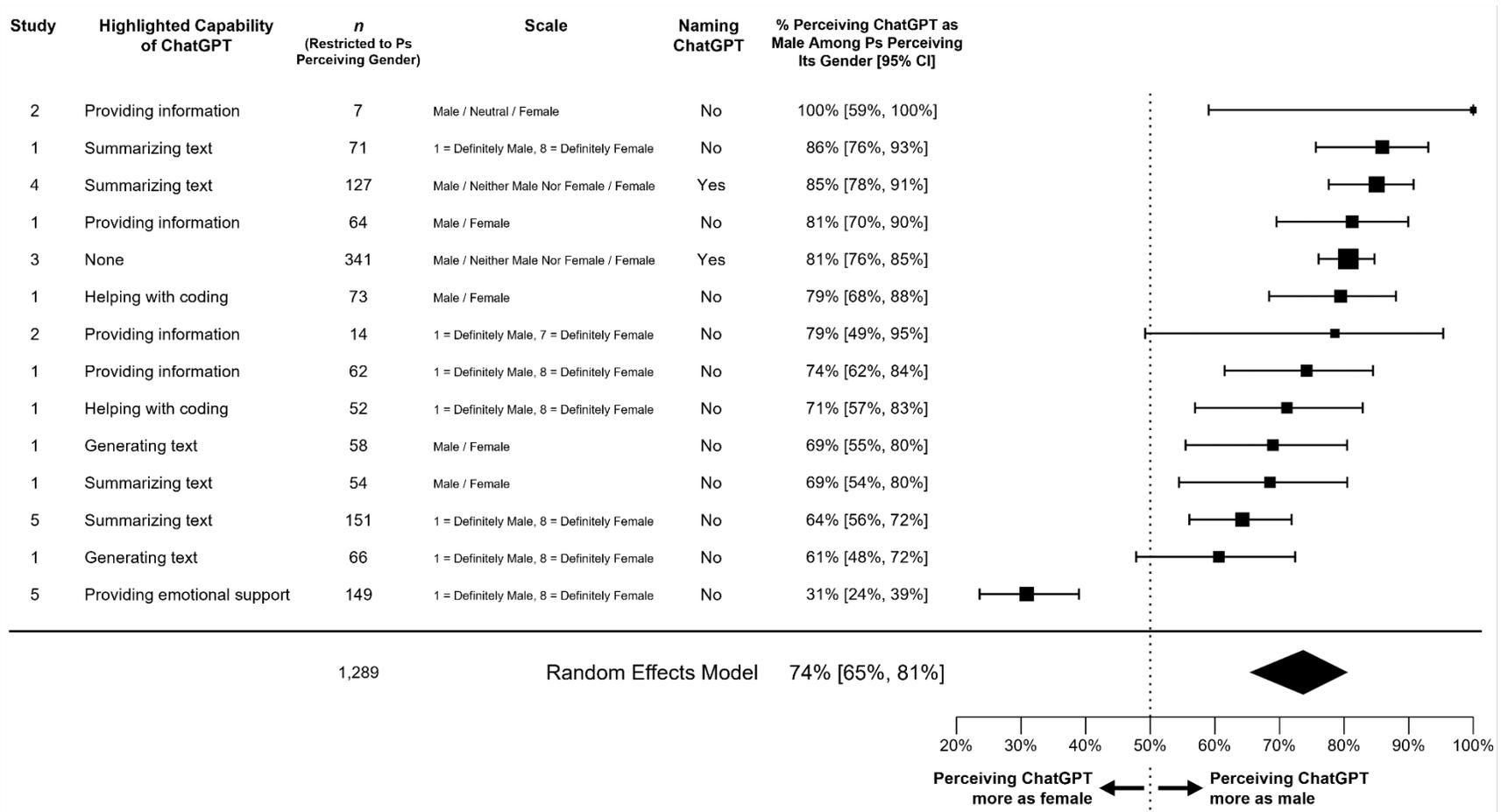

*Note.* The forest plot above shows the percentage of participants perceiving ChatGPT to be male among participants who perceived (or leaned toward perceiving) ChatGPT to be either male or female—within each subgroup of participants divided by study and study design. Within each subgroup, the male perception is more prevalent. However, the female perception is more prevalent when participants see ChatGPT provide emotional support for a user (see the last row above the horizontal line).

## General Discussion and Conclusion

Why might people perceive ChatGPT to be more male than female? Illusory stimulus *and* gender stereotype theory offer competing explanations for our findings. First, we often experience an error in face detection that leads to illusory faces being engaged by similar neural mechanisms to real faces (i.e., we see faces where there are actually none, such as in electrical outlets or tea kettles) (Palmer & Clifford, 2020). These illusory faces are devoid of any biological specification, including biological sex, emotion, or chronological age. Yet, Wardle et al. (2022) demonstrate how we perceive not only gender, but also age and emotion from these illusory stimuli. Importantly, Wardle and colleagues report a *male bias*, such that there is a strong bias to perceive these illusory faces as male. If ChatGPT is predominantly perceived to be male as our results suggest, then such a result would expand the scope of male bias currently explored in the literature, particularly when perceiving and/or judging illusory and ambiguous stimuli.

The account above, however, does not make a strong prediction about when default male perception bias should reverse; on the other hand, a gender stereotype account does. Specifically, a stereotype explanation would suggest that we hold biased beliefs about men and women. If we hold the biased belief that men are more closely associated with competence and women with warmth, and if prototypical capabilities of ChatGPT signal competence, then we may perceive ChatGPT to be male. However, in boundary cases where ChatGPT is given the opportunity to signal warmth (e.g., when providing emotional support), then we would instead perceive ChatGPT to be female. In contrast to the illusory stimulus account, the gender stereotype account makes stronger predictions about when this effect might reverse.

Previous research suggests that the conceptualization of technological assistants as females may be, in part, due to a supposed concordance between female stereotypes and passive social roles. For example, Loideain and Adams (2020) question if there is societal harm to naming Amazon's Alexa a female name, given that this technology has no recourse to refuse a request from a human. Similarly, Sutko (2020) theorizes that the association between femininity and technology reinforces gendered differences along the dimensions of docile labor, replaceable embodiment, and "artificial" intelligence. From this stream of research, one may then predict that ChatGPT, given its docility in human-machine interactions, might be conceived as female (in the same way that many existing technologies are; e.g., Apple's Siri, Amazon's Alexa, and Microsoft's Cortana). However, perhaps in part due to its ability to convey a great deal of competence, different (unfortunate) gendered stereotypes may be activated (Eagly & Mladinic, 1994), resulting in the reversed gender perception for ChatGPT.

Our findings that people in general perceive ChatGPT to be more male than female have theoretical implications. Theories that explore the psychology concerning AI and other innovative technology can take into account the perceived gender of machines and the potentially gendered interactions between humans and machines. For example, a man requesting help from Amazon's Alexa may be construed not only as a human-machine interaction but also as a *male-female* interaction. In much the same way, theorists studying evolving technology, and ChatGPT in particular, can incorporate notions of gender because, as our research demonstrates, interactions with ChatGPT may imply interactions involving a genderless partner (machine), a gendered partner (likely male but possibly female), or (a mix of) both.

An alternative perspective based on our findings may suggest that the data used to train ChatGPT, much of which is taken from the Internet and other data sources licensed to OpenAI, may have been predominantly written by men, reflecting a gender bias in the publication of written word. Should this be the case, then one may reason that a default male perception of ChatGPT simply reflects the data from which it is trained. We acknowledge that this may be a competing account of our results and invite future researchers to investigate more closely the nature, causes, and implications of perceived gender of ChatGPT.

## Pre-Registrations

Study 1: https://aspredicted.org/84Q_271

Study 2: https://aspredicted.org/HB5_SYV

Study 3: https://aspredicted.org/ CSV_7J7

Study 4: https://aspredicted.org/82D_V5V

Study 5: https://aspredicted.org/9Z2_QPR

# Appendix

**Table A1**

*Names Given to ChatGPT by Participants in Studies 3 and 4*

| Study 3 (N = 400) (Ps received minimal information about ChatGPT.) | | | Study 4 (N = 400) (Ps saw ChatGPT summarizing a news article for a user.) | | |
|---|---|---|---|---|---|
| Name | Count | % | Name | Count | % |
| Chad | 29 | 7.25 | Chad | 12 | 8 |
| Bob | 17 | 4.25 | Bob | 7 | 4.67 |
| Charlie | 10 | 2.5 | John | 5 | 3.33 |
| Alex | 9 | 2.25 | Brian | 3 | 2 |
| Chatty | 8 | 2 | Charlie | 3 | 2 |
| Adam | 7 | 1.75 | Robert | 3 | 2 |
| Charles | 7 | 1.75 | Albert | 2 | 1.33 |
| Chet | 7 | 1.75 | Barbara | 2 | 1.33 |
| John | 7 | 1.75 | Chatbot | 2 | 1.33 |
| Cathy | 6 | 1.5 | Dexter | 2 | 1.33 |
| Gary | 6 | 1.5 | Einstein | 2 | 1.33 |
| Alfred | 5 | 1.25 | Jeff | 2 | 1.33 |
| George | 5 | 1.25 | Kathy | 2 | 1.33 |
| Chip | 4 | 1 | Simon | 2 | 1.33 |
| Roger | 4 | 1 | [101 other names] | 1 (each) | 0.67 (each) |
| Sam | 4 | 1 | **Study 3 (Continued)** | | |
| Steve | 4 | 1 | Buddy | 2 | 0.5 |
| Alan | 3 | 0.75 | Chappie | 2 | 0.5 |
| Carl | 3 | 0.75 | Chat | 2 | 0.5 |
| Greg | 3 | 0.75 | Chaz | 2 | 0.5 |
| Hal | 3 | 0.75 | Chester | 2 | 0.5 |
| Henry | 3 | 0.75 | Dave | 2 | 0.5 |
| Joe | 3 | 0.75 | Einstein | 2 | 0.5 |
| Kevin | 3 | 0.75 | Frank | 2 | 0.5 |
| Pat | 3 | 0.75 | Harold | 2 | 0.5 |
| Pete | 3 | 0.75 | Jane | 2 | 0.5 |
| Rob | 3 | 0.75 | Jason | 2 | 0.5 |
| Stella | 3 | 0.75 | Jerry | 2 | 0.5 |
| Tom | 3 | 0.75 | Jim | 2 | 0.5 |
| Al | 2 | 0.5 | Karen | 2 | 0.5 |
| Andy | 2 | 0.5 | Kathy | 2 | 0.5 |
| Anna | 2 | 0.5 | Larry | 2 | 0.5 |
| Arnold | 2 | 0.5 | Rick | 2 | 0.5 |
| Boris | 2 | 0.5 | Stupid | 2 | 0.5 |
| Bot | 2 | 0.5 | Tony | 2 | 0.5 |
| Brian | 2 | 0.5 | [173 other names] | 1 (each) | 0.25 (each) |